# Enhanced lasing action and spontaneous emission in periodic plasmonic nanohole arrays.

Bryson Krause,[1] Minh Pham,[2] Hoang Luong,[2] Tho Nguyen,[2] and Thang Hoang[1, a)]

[1]*Department of Physics and Material Science, University of Memphis, Memphis, TN 38152*
[2]*Department of Physics & Astronomy, University of Georgia, Athens, GA 30602*
[a)]*Author to whom correspondence should be addressed. Electronic mail: tbhoang@memphis.edu*

Periodic arrays of air nanoholes in thin metal films that support surface plasmon resonances can provide an alternative approach for boosting the light-matter interactions at the nanoscale. Indeed, nanohole arrays have garnered great interest in recent years for their use in biosensing, light emission enhancement and spectroscopy. However, the large-scale use of nanohole arrays in emerging technology requires new low-cost fabrication techniques. Here, we demonstrate a simple technique to fabricate nanohole arrays and examine their photonic applications. In contrast to the complicated and most commonly used electron beam lithography technique, hexagonal arrays of nanoholes are fabricated by using a simple combination of shadowing nanosphere lithography technique and electron beam deposition. These arrays are shown to offer enhancements in the lasing emission of an organic dye liquid gain medium with a quality factor above 150. Additionally, a 7-fold increase in Purcell factor is observed for CdSe quantum dot-integrated nanohole arrays.

*Keywords: Plasmonics, nanohole arrays, nanolasers, Purcell enhancement*

## I. INTRODUCTION

Periodic lattices of plasmonic nanostructures are well established in research and industry as a powerful tool for ultra-sensitive detection and high spectral resolution at the nanoscale level.[1,2] Nanostructured plasmonic metasurfaces such as nanoparticle or nanohole arrays have been exploited for a wide range of applications including biosensing, light harvesting and optofluidic devices.[3-10] Having been studied extensively over the last two decades, the attributes of many types and configurations of nanostructured arrays are thoroughly documented.[11,12] One of the biggest present challenges lies in producibility and utilization, which is largely dependent on a low cost and ease of fabrication. There are many well-understood methods of fabrication which yield a variety of nanoarrays.[13-16] The ideal nanoarray involves a fabrication process that is simple, cheap and precise, though many of the nanoarray fabrication techniques used in practice are cumbersome and expensive.[17,18] In this paper, a low-cost alternative for fabricating periodic nanohole arrays (PNAs) onto a glass substrate is described. In contrast to the conventional electron beam lithography technique, large surface area of PNAs can be rapidly fabricated. More importantly, fabricated PNAs are shown to exhibit a strong surface lattice plasmon resonance which hints at useful applications in photonics and optoelectronics.

Periodic nanohole arrays have the practical ability to strongly confine light via surface lattice plasmons, thus offering a way forward with controlling the light-matter interaction at an extended length scale of many lattice constants.[19,20] The design parameters for PNAs that most influence their plasmonic properties are the periodicity and size of the nanoholes. Control of these parameters and the directional radiation ability of nanohole arrays is crucial for their practical application. The use of liquid gain media on a nanoparticle array has been previously demonstrated to achieve a dynamically tunable lasing response in a plasmonic nanoparticle array.[21] Similarly, a liquid gain fluorescent dye medium is shown to induce an efficient lasing response in the PNAs discussed in this study.

Because of the high photonic density of states provided by plasmonic arrays in a small volume, it is possible to drastically increase the spontaneous emission rate of coupled quantum systems at particular wavelengths.[22] This effect was first observed in the 1940s and is named the Purcell effect after its discoverer, Edward M. Purcell.[23] The strength of this phenomenon in a given system can be quantified by the so-called Purcell factor (PF), which is related to the decrease in measured photon decay time of an emitter due to its surrounding environment. Faster decay rates associated with an increased fluorescent intensity can yield more efficient and effective electronic devices that utilize plasmonic systems.[22] The PF can be determined experimentally from the ratio of the lifetimes of a quantum system both in and out of a nanocavity environment, and such an increase in the PF is demonstrated in PNAs paired with quantum dots (QDs) in this study.

## II. METHODS

### A. Nanohole array fabrication

Reactive ion etching (RIE) is a dry etching process by which material deposited onto a wafer is removed by chemically reactive plasma, which is produced by an electromagnetic field in a low-pressure environment. For the fabrication of nanohole arrays used in this study, a colloidal polystyrene (PS) nanosphere monolayer was placed onto a glass substrate through the air-water interface method, as described previously.[24]

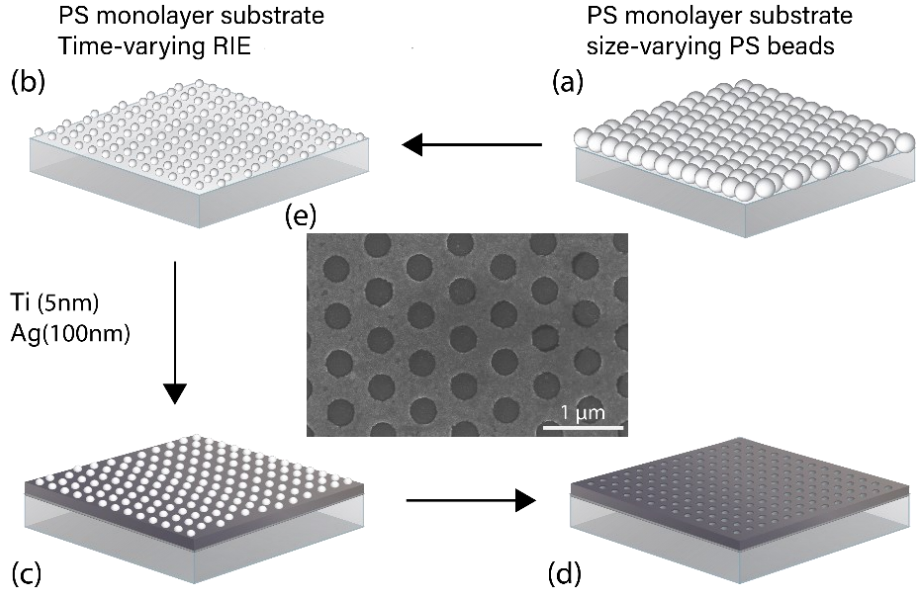

FIG. 1. Nanohole array fabrication steps. (a) PS beads are deposited in a single layer onto a glass substrate. (b) PS beads are then subjected to RIE to decrease bead size, thus reducing the diameter of nanoholes. (c) 5-nm Ti and 100-nm Ag are deposited onto the PS/glass configuration. (d) The PS layer is then lifted from the assembly leaving a periodic hexagonal nanohole array. (e) SEM image of a nanohole array.

To summarize this method, the PS beads are first dissolved in a solution of water and ethanol, which is dropped slowly in small increments onto a glass petri dish containing a small amount of water. The surface area of the PNA, which is typically a few centimeters large in this present study, is determined by the size of the glass substrate. After some time, the PS beads form a hexagonally close packed monolayer on the water-air interface, and a Teflon ring is placed around the surface to protect the monolayer from adhesion to the side of the petri dish. The water level is then raised, and a glass substrate is carefully slid beneath the PS monolayer film. The water then gets pumped out, and the PS monolayer film is lowered onto the glass substrate. Once the monolayer is lowered completely, the PS beads are subjected to RIE, effectively decreasing the bead size. After RIE, a 5-nm layer of Ti followed by a 100-nm Ag layer is deposited via the electron-beam evaporation onto the glass/PS array at the base pressure of $10^{-6}$ mbar. The Ti layer acts as a bonding agent between the glass substrate and the Ag layer. After layer deposition, the PS beads are stripped away from the assembly with Scotch tape, leaving behind a periodic hexagonal nanohole array in Ag. The controllable parameters in this process (Fig. 1 (a) - (d)) are the lattice spacing and nanohole size, which can be adjusted through the size of PS beads and RIE exposure time, respectively. The nanohole array samples presented here were made to have an inter-particle spacing of 500-nm, with radii of approximately 150-nm corresponding to a respective RIE time of 300 s. Figure 1(e) shows the scanning electron microscope (SEM) image of an PNAs sample used for this work.

## B. Sample characterization

In order to determine their resonant modes, each nanoarray was mounted onto a controllable turntable, and an angle-dependent transmission scan was taken for 45 degrees about normal incidence. Figure 2(a) shows the peak resonances at normal incidence of a 500-nm lattice array resulting from several RIE etching time of 300 s. The spectrum contains well-defined transition maximum at ~ 850-nm due to the Bragg diffraction mode and transmission minima at ~ 660-nm due to the Wood - Rayleigh anomaly modes at Ag/air and Ag/glass interfaces, respectively.[25,26] In this situation the light that couples to these resonant modes is pointing in the direction perpendicular to the sample surface. This is very important, because the photons emitted by an embedded emissive nanomaterial will be guided by the directional resonance. At the 660-nm transmission minimum the full-width-at-half -maximum (FWHM) of this transmission curve was determined to be around 40-nm which results in an approximate quality factor of 16.

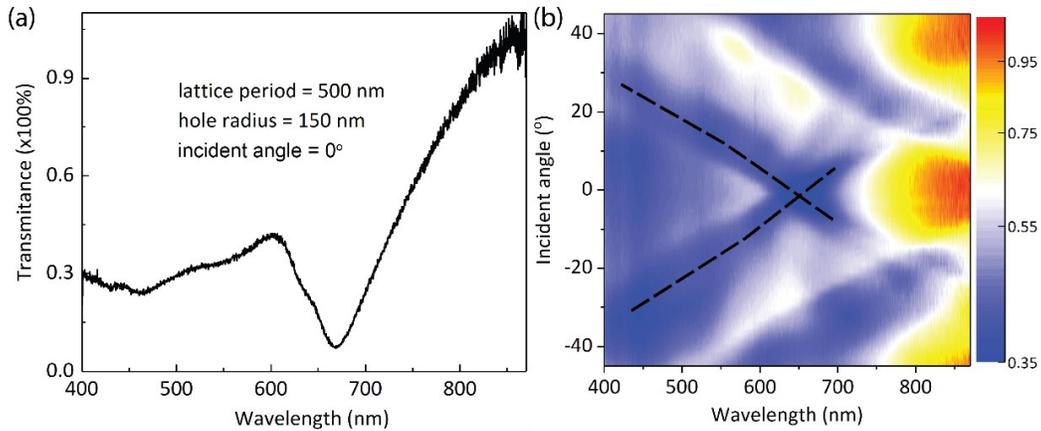

FIG. 2. Optical characterizations of PNAs. (a) Transmission curves of PNAs for varying nanohole sizes at the normal incident angle. (b) Angle resolved transmission of an PNA. Dashed lines indicate lattice plasmon resonances that follows the (0, ±1) Wood – Rayleigh modes.

Figure 2(b) shows a false-color image which was constructed by transmission spectra collected as a function of incident angles. The dashed curves indicate the (0, ±1) Wood - Rayleigh anomaly Ag/air mode position as a function of the incident angle. This observation is similar to the resonant modes that were observed by several previous reports.[27-30] In the following sections, we will be interested in this particular resonant mode and the resonant wavelength at the normal incident angle will be used as a reference for the selection of appropriate fluorescent emitters.

## C. Active materials integration and optical measurements

In order to probe the lasing enhancement behavior of the PNAs, several different configurations of active medium on glass slide and on nanoarray were prepared. To stimulate lasing, an organic fluorescent DCM dye [*4-(Dicyanomethylene)-2-methyl-6-(4-dimethylaminostyryl)-4H-pyran*] suspended in DMSO [*Dimethyl Sulfoxide*] was utilized as the active medium. The spectrum of DCM dye showed a resonant emission at around 663-nm, which nearly overlaps with the transmission minima at ~ 660-nm of the PNA. A range of concentrations of dye in DMSO was mixed in a vortexer and sonicated for 5 minutes, and 15 μL of each concentration was transferred with a pipette onto separate 1 cm square pieces of either slide glass or PNA to be used as a liquid gain medium for lasing. In an ideal configuration, the dye molecules would rest in the nanoholes. However, as this could not be achieved with the methods used here, the dye molecules most likely remained suspended above the nanoholes in the liquid medium.

Spontaneous emission enhancement by the PNAs can also be demonstrated by observing the shortened decay time of QDs integrated into the nanoarrays. Several 1 $cm^2$ pieces of the 500-nm lattice, 150-nm radius hole PNA were covered with a diffuse concentration (0.01mg/mL) of CdSe QDs (Sigma Aldrich) suspended in toluene. The sample was spin-coated for 30 seconds at 1500 revolutions per minute. This configuration was washed clean with deionized water and flushed with clean nitrogen gas. Spectral analysis of the emission of the CdSe QDs confirmed the manufacturer's specified peak emission value of 655-nm, which is close to the PNA resonance at around 663-nm.

All optical measurements were taken at standard atmospheric pressure and at ambient temperature. Photoluminescence (PL) signals were captured with a CCD (charge-coupled device) camera and analyzed with a Horiba iHR550 spectrometer. For the decay time measurements of the quantum dots, a time-correlated single photon counting setup was used. An ultrafast Coherent laser at 475-nm (80MHz, 150fs) was used to excite the QDs through a 10X objective lens. The PL emission from the QDs was collected by the same objective lens and sent through the spectrometer for either spectral analysis or through a side exit to be guided into a fast-timing avalanche photodiode for temporal analysis. The signals collected by the photodiode were analyzed by a single photon counting module (PicoQuant PicoHarp 300). For the lasing experiment, DCM dye was excited by another Coherent laser at 515-nm (2 kHz, 2 ns pulse width) by using a 5x objective lens at a normal incident angle. The PL emission from the dye was collected by the same lens and then dispersed and analyzed by the spectrometer and CCD camera. It is also important to note that for the objective lenses used in our experiments, the excitation and collection spot size was approximately 100 μm in diameter, which covered an area of many lattice periods (Figs. 3(a) and 4(a)).

### III. RESULTS AND DISCUSSION

In this section, we demonstrate the lasing action as well as enhanced spontaneous emission of quantum emitters by integrating them with the PNAs. Nanohole arrays with surface lattice plasmon resonances at around 665-nm were used. Because of this, organic DCM dye and colloidal CdSe QDs, which have a similar emission wavelength, were chosen as active materials. It is worth noting that several similar experiments have previously demonstrated such applications of different nanoparticle arrays, and our current purpose is to show that with an alternative simple fabrication technique, we can obtain similar results for potential useful applications in nanophotonics and optoelectronics at a lower cost and large-scale production.[21,31]

### A. Lasing enhancement

We first attempt to demonstrate potential applications of nanohole arrays by taking advantage of the surface lattice plasmon resonance to enhance lasing action of embedded organic DCM dye. This creates opportunities for developing plasmonic chips with integrated directional, efficient coherent light sources.

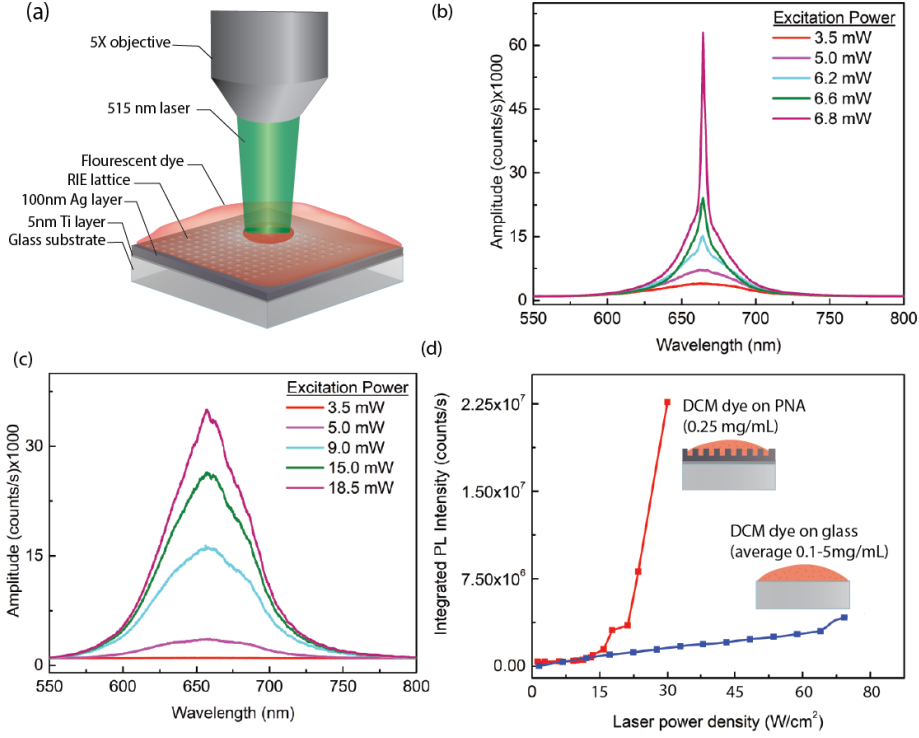

FIG. 3. Enhanced lasing emission by PNAs. (a) Experimental setup consisting of a 2 kHz at 515-nm and 2 ns-pulse laser transmitted through a 5x objective onto a DCM-dye drop placed onto a 300-nm diameter nanohole array with a 500-nm lattice spacing. Power dependent PL spectra of 0.5 mg/mL dye on (b) nanoarray and (c) on glass. (d) Comparison between the integrated PL peak intensities of 0.25 mg/mL DCM dye on nanohole array and the average peak intensity of DCM dye concentrations on glass, ranging from 0.1 to 5 mg/mL. Square dots are data points and solid lines between dots are guides to the eye.

To investigate the lasing enhancement offered by the PNAs, the PL spectra were recorded for both DCM dye-covered glass and dye-covered 500-nm lattice (150-nm hole radius) at equal intervals starting at a power density of 11.3 $\frac{W}{cm^2}$ (or at 3.5 mW incident power) until CCD reading saturation, which occurred at around 59.7 $\frac{W}{cm^2}$ for dye on glass, or until lasing was achieved for dye in PNA. Figures 3 (b) and (c) show the PL spectra of the DCM dye under various laser excitation densities for a dye concentration of 0.5 mg/mL for both cases. It is also observed that the dye covered nanoarray showed a more prominent lasing response at a lower excitation power than the dye/glass for multiple concentrations of dye. This lasing response began at a threshold of about 18.9 $\frac{W}{cm^2}$, for dye on nanoarray, as indicated in the nonlinearity in the integrated peak intensity plot shown in Fig. 3(d), which shows the compared PL responses between nanoarray and glass with DCM dye. At a power density of 22.0 $\frac{W}{cm^2}$, a plasmonic supported lasing mode was fully evident in the nanohole array with a FWHM of 4.2-nm and a corresponding quality $Q = \frac{\lambda}{\Delta\lambda}$ greater than 150 (Fig. 3 (b), which is comparable with values presented in other studies using fluorescent emitters combined with comparable nanostructure arrays.[21,32] For the same dye concentration on glass, even at high excitation densities the spectrum exhibits a broad PL band with a FWHM of 50-nm, indicating only spontaneous emission (Fig. 3(c)). It should be noted that the emission intensity of DCM dye on glass represented in Fig. 3 (d) is averaged for dye concentrations varying from 0.1 to 5mg/mL. For the larger concentration dyes on glass, lasing was evident at a high-power density, however, for the same dye placed on PNA lasing occurred at a much lower power. At a concentration of 1mg/mL, for example, dye on glass exhibited lasing at a power density of about 69.0 $\frac{W}{cm^2}$ while the dye on PNA configuration began to lase at 19.7 $\frac{W}{cm^2}$ (result not plotted here), thus demonstrating a lasing threshold reduction factor of 3.5 for dye on PNA.

Such a lasing emission in PNAs is of potential applications for on-chip coherent light sources, which can transcend the physical limitations imposed on conventional lasers by the optical diffraction limit.[33] Additionally, the use of organic dye as a liquid gain medium has been shown to be a promising method for providing a means of dynamic tunability in these type of plasmonic nanostructures.[20]

**B. Enhanced spontaneous emission.**

Novel nanomaterials, such as semiconductor quantum dots,[34,35] nanowires[36,37] or 2D dichalcogenides[38,39] offer properties and applications that do not exist in their macroscopic counterparts. However, current photonic devices based on these nanomaterials are limited by their intrinsic properties, including slow temporal emission (10-20 ns), low efficiency (10-15 %), and non-directional spatial emission. Integrating these nanomaterials with plasmonic nanocavities, which provide high photonic density of states, will significantly enhance their overall efficiency. As a result, such a device can significantly boost overall device efficiency to more than 50%[40,41] and operating frequency up to 90 GHz[41] yielding promise for photonic applications ranging from sensing and information processing to photovoltaics. Even though many plasmonic devices have previously demonstrated to offer significant enhanced optical properties such as PL intensity, spontaneous emission rate or surface enhanced Raman signal, our intent here is to demonstrate that PNAs could also provide an alternative path for enhancing these properties.

In our experiment, the PL and decay time measurements of a 500 nm-lattice, 150-nm radius hole PNA covered with CdSe QDs were taken and compared to that of the exact same CdSe QDs on a glass slide as control. The samples were excited with a power density of 3 $\frac{W}{cm^2}$ by a pulsed, frequency doubled laser (Coherent Chameleon Ultra II, 80 MHz, 150 fs) tuned to 475-nm through a 10x objective (Fig. 4(a)). Figure 4(b) exhibits the brighter PL emission of the QDs on the nanohole array (red) than that on a glass slide (blue). A transmission curve of the nanohole array, which indicates the surface plasmon resonance at exact same wavelength with the QD emission, is also plotted. The decay curves were acquired by a time-correlated single photon counting module and plotted as the coincident count vs. time for each sample, as shown in Fig. 4 (c). While the intrinsic decay time of QDs on a glass slide exhibits a single decay component, the QDs in the PNA appear to show a fast and a slow decay channel. From the fitting of the average of these plots spanning from the decay measured at 645 to 700-nm, a decay lifetime of 9.6 ns was determined for CdSe QDs on glass. For the fast and slow decay components of CdSe QDs on PNA, 1.35 ns and 6.0 ns were obtained at the peak transmission wavelength of the PNA. The faster decay time of the QDs in the array associated with an enhanced PL intensity therefore indicates an enhanced spontaneous emission PF of approximately 7. Several previous reports of the PF for nanoparticle arrays or nanohole arrays also show similar results.[42,43] Indeed, the observed PL intensity and decay time in the array involved both QDs that coupled to the surface plasmon mode and QDs those did not couple to the lattice and the slow decay component observed from QDs in the array is related to the non-coupled QDs. Figure 4(d) shows the decay times of QD on glass and PNA (the fast component) which were sampled at a 5-nm interval from 645-nm to 700-nm, following the bounds of each sample's PL curves. It is important to note that the QDs on the PNA exhibit a fast decay time across its emission spectrum, indicating the efficient coupling with the PNA's plasmon resonance.

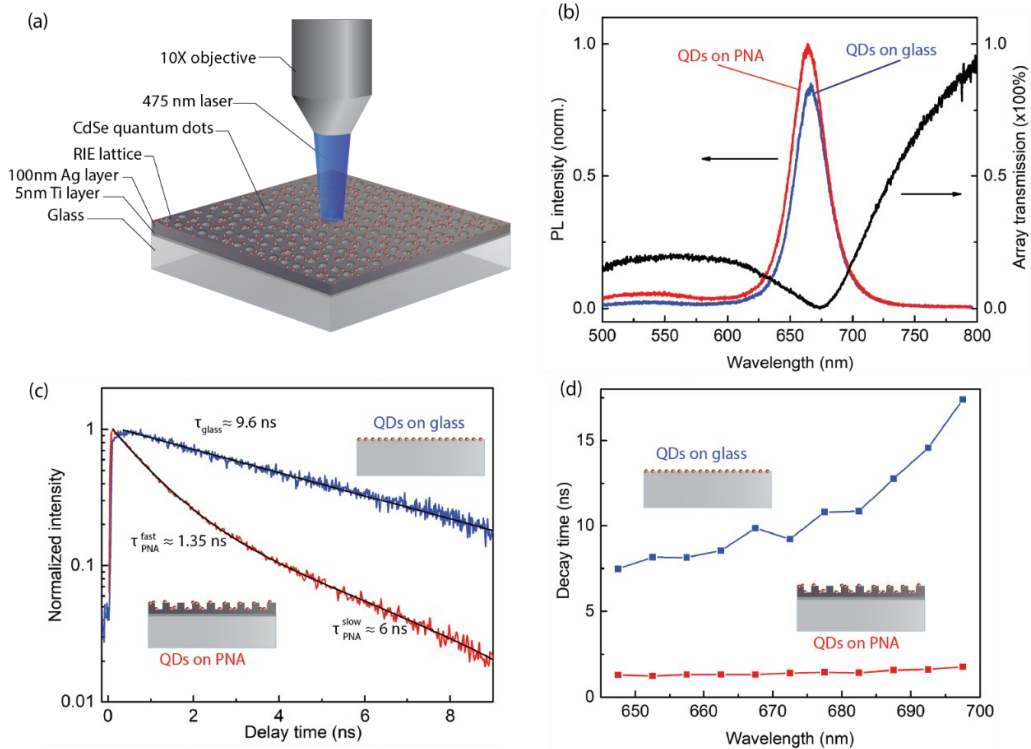

FIG. 4. Enhanced spontaneous emission by PNAs. (a) Schematic of experimental setup for measuring CdSe QD-covered PNA. (b) Comparison between the PNA's transmission (black), PL emission of CdSe QDs on glass (blue) and on PNA (red). (c) Average measured decay curves of CdSe QDs on glass (black) and on a 500-nm lattice PNA (red). Black curves are fits to the data. (d) Wavelength dependent decay times of QDs on glass and on the array (fast component). Square dots are data points while solid lines between dots are guides to the eye.

The results of the lasing and spontaneous emission enhancement measurements described in this work point to several practical applications such as ultrafast light-emitting diodes, optical sensing or quantum information processing systems for this simple and inexpensive periodic PNA design. The control of a photonic material's spontaneous emission rate is key in the efficient generation of photons, and QD-based devices need to be designed with this in mind.

## IV. CONCLUSION

In summary, PNAs were designed and constructed by using a simple combination of shadowing nanosphere lithography technique and electron beam deposition. In contrast to the conventional electron beam lithography approach, the technique presented in this work is simple, cost-effective, and more importantly, large area of high-quality metasurfaces can be rapidly fabricated. These arrays were demonstrated to have enhanced lasing capabilities when coupled with an organic dye in a liquid gain medium. Furthermore, the use of CdSe quantum dots on the arrays led to an enhanced spontaneous emission rate with a Purcell factor of approximately 7. These simple nanohole arrays can therefore offer further prospect in the industrialization of nanoarrays for use in various realms of emerging technologies such as sensing, ultrafast or on-chip coherent light sources.

## ACKNOWLEDGMENTS

This work is supported by the National Science Foundation (NSF) (Grant # DMR-1709612). TH acknowledges the support from the FedEx Institute of Technology at the University of Memphis. UGA authors acknowledge the support from Thomas Jefferson Fund (Grant # RFACE0001080201).